# Strongly baryon-dominated disk galaxies at the peak of galaxy formation ten billion years ago[†]

R. Genzel[1,2*], N.M. Förster Schreiber[1*], H. Übler[1], P. Lang[1], T. Naab[3], R. Bender[4,1], L.J. Tacconi[1], E. Wisnioski[1], S. Wuyts[1,5], T. Alexander[6], A. Beifiori[4,1], S. Belli[1], G. Brammer[7], A. Burkert[3,1], C.M. Carollo[8], J. Chan[1], R. Davies[1], M. Fossati[1,4], A. Galametz[1,4], S. Genel[9], O. Gerhard[1], D. Lutz[1], J.T. Mendel[1,4], I. Momcheva[10], E.J. Nelson[1,10], A. Renzini[11], R. Saglia[1,4], A. Sternberg[12], S. Tacchella[8], K. Tadaki[1] & D. Wilman[4,1]

[1]*Max-Planck-Institut für extraterrestrische Physik (MPE), Giessenbachstr.1, 85748 Garching, Germany*
    (*genzel@mpe.mpg.de, forster@mpe.mpg.de*)
[2]*Departments of Physics and Astronomy, University of California, 94720 Berkeley, USA*
[3]*Max-Planck Institute for Astrophysics, Karl Schwarzschildstrasse 1, D-85748 Garching, Germany*
[4]*Universitäts-Sternwarte Ludwig-Maximilians-Universität (USM), Scheinerstr. 1, München, D-81679, Germany*
[5]*Department of Physics, University of Bath, Claverton Down, Bath, BA2 7AY, United Kingdom*
[6]*Dept of Particle Physics & Astrophysics, Faculty of Physics, The Weizmann Institute of Science, POB 26, Rehovot 76100, Israel*
[7]*Space Telescope Science Institute, Baltimore, MD 21218, USA*
[8]*Institute of Astronomy, Department of Physics, Eidgenössische Technische Hochschule ETH Zürich, CH-8093, Switzerland*
[9]*Center for Computational Astrophysics, 160 Fifth Avenue, New York, NY 10010, USA*
[10]*Department of Astronomy, Yale University, 260 Whitney Avenue, New Haven, CT 06511, USA*
[11]*Osservatorio Astronomico di Padova, Vicolo dell'Osservatorio 5, Padova, I-35122, Italy*
[12]*School of Physics and Astronomy, Tel Aviv University, Tel Aviv 69978, Israel*

**In the cold dark matter cosmology, the baryonic components of galaxies - stars and gas - are thought to be mixed with and embedded in non-baryonic and non-relativistic dark matter, which dominates the total mass of the galaxy and its dark matter halo[1]. In the local Universe, the mass of dark matter within a galactic disk increases with disk radius, becoming appreciable and then dominant in the outer, baryonic regions of the disks of star-forming galaxies. This results in rotation velocities of the visible matter within the disk that are constant or increasing with disk radius – a hallmark of the dark matter model[2]. Comparison between the dynamical mass and the sum of stellar and cold-gas mass at the peak epoch of galaxy formation ten billion years ago, inferred from ancillary data, suggest high baryon factions in the inner, star-forming regions of the disks[3-6]. Although this implied baryon fraction may be larger than in the local Universe, the systematic uncertainties (stellar initial mass function, calibration of gas masses) render such comparisons inconclusive in terms of the mass of dark matter[7]. Here we report rotation curves for the outer disks of six massive star-forming galaxies, and find that the rotation velocities are not constant, but decrease with radius. We propose that this trend arises because of a combination of two main factors: first, a large fraction of the massive, high-redshift galaxy population was strongly baryon dominated, with dark matter playing a smaller part than in the local Universe; and second, the large velocity dispersion in high-redshift disks introduces a substantial pressure term that leads to a decrease in rotation velocity with increasing radius. The effect of both factors appears to increase with redshift. Qualitatively, the observations suggest that baryons in the early Universe efficiently condensed at the centres of dark matter halos when gas fractions were high, and dark matter was less concentrated.**

Over the last few years, there have been significant studies of ionised gas dynamics of redshift 0.6-2.6 star-forming galaxies. These were drawn from mass-selected parent samples in cosmological deep fields, and from imaging and grism surveys with the Hubble Space telescope [3, 5, 8-13], with well characterized properties (see Methods). We obtained deep (5-22 h integration), imaging spectroscopy of the Hα emission line, with the near-infrared integral field spectrometers SINFONI and KMOS on the Very Large Telescope of the European Southern Observatory, as part of our "SINS/zC-SINF" and "KMOS[3D]" surveys (henceforth 'IFS-samples' [3,8]). From the data we extract Hα rotation curves, rotation velocities as a function of galactic radius, for several hundred star-forming galaxies (Methods). Rotation curves give valuable constraints on the baryonic and dark matter mass distributions in massive star-forming disks at the peak of cosmic galaxy formation 10 billion years ago, largely independent of assumptions on star-formation histories and stellar mass functions.

Figure 1 shows the angular distributions of stellar mass/light, integrated Hα intensity, Hα velocity and dispersion, together with cuts of the latter two along the kinematic major axis, for six of our best star-forming galaxies. We selected galaxies for deep spectroscopy from the IFS-samples to have large stellar masses ($\log(M_*/M_\odot)$~10.6-11.1), to not participate in a merger, to have rotationally dominated kinematics, and to have large half-light radii ($R_{1/2}$~4-9 kpc), such that the disk penetrates far into the dark matter halo.

The extracted velocity and velocity dispersion fields in all cases demonstrate that the sources are rotationally supported[8] (Methods). The ratio of peak rotation velocity to the amplitude of random motions, as estimated from the velocity dispersion in the outer disks, is between 4 and 9, lower than in present-day disks ($v_{rot}/\sigma_0$~10-20), and in excellent agreement with other observations[3,8]. This means that 'turbulent' motions contribute significantly to the energy balance[14].

The ***most surprising*** new result is that the projected rotation velocities along the kinematic major axis (panels (c) in Figure 1) reach a maximum value $|v_{max}|$ at $R_{max}$ and decrease further out, symmetrically on either side of the galaxy centre. Averaging the two sides of the galaxy further improves the signal-to-noise ratio (Figure 2). The six rotation curves drop to $v(R_{out})/v_{max}$~0.3 to 0.9 at the outermost radius sampled. Falling rotation curves have previously been detected at low redshift in some compact, high surface density, or strongly bulged disks (such as Andromeda, Figure 2), although these are rare and drops are modest, to $v/v_{max}$~0.8-0.95 [2,15-17].

How common are the falling rotation curves at high redshift? Reference 18 provides a co-added rotation curve of 97 rotationally supported z~0.6-2.6 isolated disks from the same IFS samples (***excluding*** the six above), providing a representative and fairly unbiased sample of the redshift-stellar mass-star formation rate, parameter space for $\log(M_*/M_\odot)$>9.7 star-forming galaxies (see Methods and Extended Data Figure 1). Star-forming galaxies over the entire mass range of the parent sample enter the stack (Extended Data Figure 1). Within the uncertainties, the stack confirms the results presented here for individual sources, and implies that falling rotation curves at z=0.7-2.6 are common. The uncertainties of individual velocity measurements in the faint outer disks are substantial, such that the significance of the velocity drops in each individual data point is ≤3.5 rms. When all data points at >$R_{max}$ are considered together the statistical significance for a non-flat, falling rotation curve becomes compelling (6-10 rms).

We investigated the possibility that the falling rotation curves are an artefact, caused by warping of the disks, radial streaming along galactic bars, radial changes in the direction of the kinematic axis, tidal interactions with nearby satellites, non-equilibrium motions caused by variations in the amount and/or the direction of the baryonic accretion (Methods). We find interacting low mass satellites in three of our six sources and evidence for some tidal stripping in one, but the rotation curve is symmetric, even near the satellite. We do see strong radial streaming confined to the nuclear region in one galaxy. Warps are expected because of the non-planar accretion of gas from the intergalactic web, but also less likely at high-z than in the low-z Universe to be stable, because of the large, isotropic velocity dispersions. The point-symmetric, falling rotation curves in Figure 1 argue against strong warping. We also find no evidence in the two dimensional residual maps (data minus model) for radial variations in the line of nodes, as a result of interactions and variations in the angular momentum of the incoming gas (Extended Data Figure 6). Four galaxies have massive bulges (B/T>0.3), which likely will accentuate centrally peaked rotation curves. Keeping in mind the important effects of non-equilibrium dynamics in the early phases of galaxy formation, the prevalence of point-symmetric, smooth rotation curves in all six cases suggests that these are intrinsic properties of the galaxies.

We compare the final average of all seven rotation curves with an average rotation curve of local massive disks[19], the curves of the Milky Way[20] and M31[17], and the theoretical curve of a thin, purely baryonic, exponential 'Freeman' disk[21] in panel (b) of Figure 2. All local rotation curves are above the Freeman model, and thus require additional (dark) matter in various amounts. This is not the case for the average high-redshift curve. It is consistent with a pure baryonic disk to $R\sim 1.8\ R_{1/2}$, and falls below it further out.

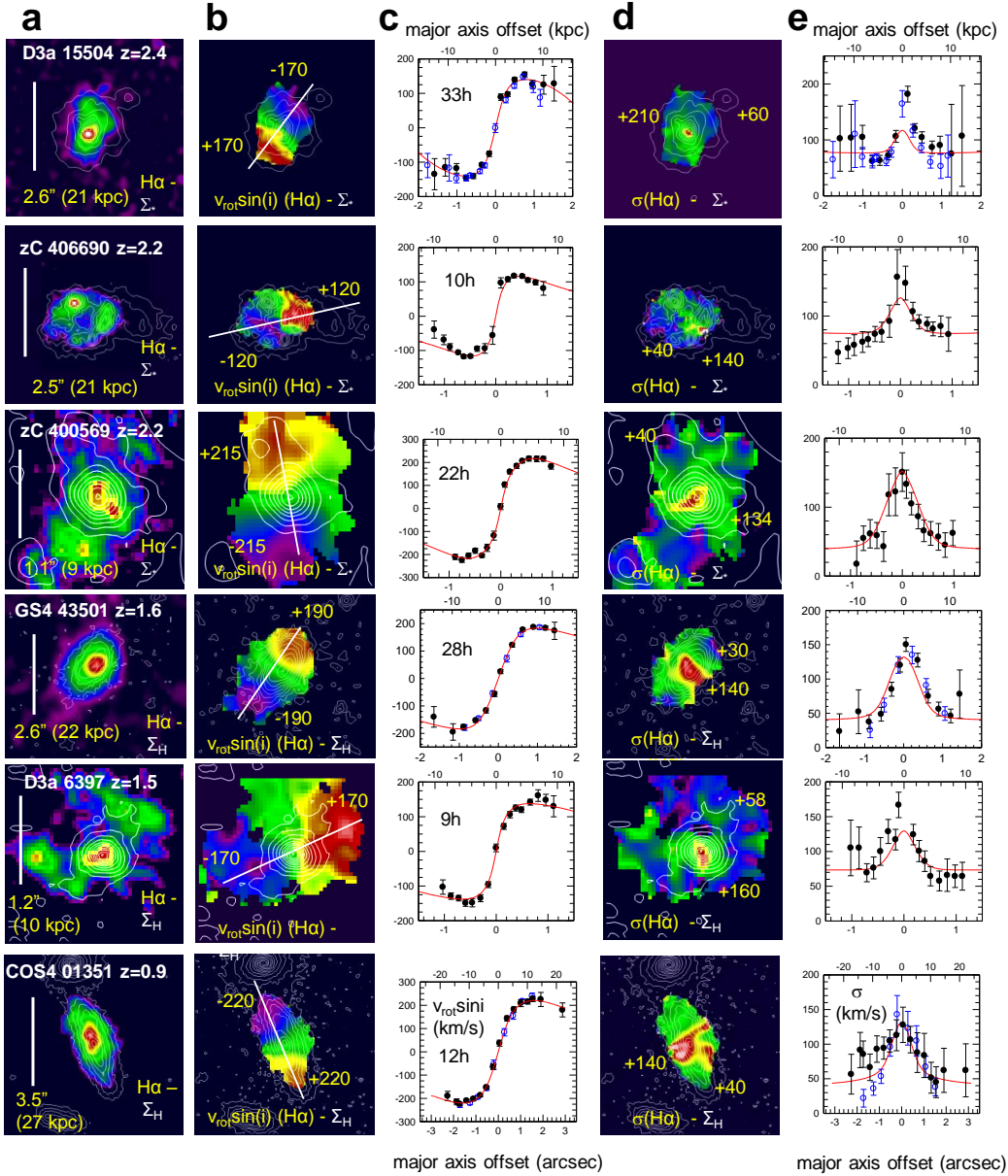

**Figure 1. Hα gas dynamics from KMOS and SINFONI in six massive star-forming galaxies.** The galaxies have redshifts between z=0.9 and 2.4. KMOS provides seeing-limited data (FWHM ~0.6"), and SINFONI allows both seeing-limited, and adaptive optics assisted observations (FWHM ~0.2"). For each galaxy column (**a**) shows the distribution of the integrated Hα line surface brightness (with a linear colour scale), superposed on white contours of the stellar surface density ($\Sigma_*$) or the H-band continuum surface brightness ($\Sigma_H$) (square root scaling). The vertical white bar denotes the physical size scale. Column (**b**) gives the velocity map (colour scale, with extreme values indicated in km/s), superposed on $\Sigma_*$ or $\Sigma_H$ contours (white lines, square-root scaling), derived from fitting a Gaussian line profile to the Hα data in each pixel (0.05"). The top four galaxies have FWHM ~0.25" (2 kpc), the bottom two sources have FWHM ~0.55-0.67" (5 kpc). Column (**c**) shows the extracted line centroids (and ±1rms uncertainties) along the kinematic major axis (white line in (**b**)). For COS4 01351 and GS4 53501 we show both SINFONI (black filled circles) and KMOS (open blue circles) data sets, for D3a 15504 we show SINFONI data sets at 0.2" (black) and 0.5" (blue) resolution. Red continuous lines denote the best-fit dynamical model, constructed from a combination of a central compact bulge, an exponential disk and an NFW halo without adiabatic contraction, with a concentration of c=4 at z~2 and c=6.5 at z~1. For the modelling of the disk rotation, we also take into account the asymmetric drift correction inferred from the velocity dispersion curves ((**d**)-(**e**))[14]. Columns (**d**) and (**e**) give the two-dimensional and major-axis velocity dispersion distributions (colour) inferred from the Gaussian fits (after removal of the instrumental response, $\sigma_{instr}$~37 km/s at z~0.85 and 2.2 and ~45 km/s at z=1.5), superposed on $\Sigma_*$, or $\Sigma_H$ maps (white contours); the numbers in (**d**) indicate the minimum and maximum velocity dispersion. All physical units are based on a concordance, flat $\Lambda CDM$ cosmology with $\Omega_m$=0.3, $\Omega_{baryon}/\Omega_m$=0.17, $H_0$=70 km s$^{-1}$ Mpc$^{-1}$.

Since high-redshift disks exhibit large random motions, the equation of hydrostatic equilibrium of the disk contains a radial pressure gradient, which results in slowing down the rotation velocity ('asymmetric drift'[14], Methods). If we apply this correction and also allow for the resulting significant thickness of the disk, the rotation curve indeed drops rapidly with radius, as long as $\sigma_0$ stays approximately constant. Figure 2 shows a $v_{rot}/\sigma_0\sim 5$ galaxy, which provides an excellent match to the average observations.

Our analysis leaves little space for dark matter in the outer disks (and inner halos) of massive, high-redshift star-forming galaxies. This conclusion is consistent with the earlier, but less secure, analysis of the 'inner disk dynamics'[3-6] (Methods). We quantify our conclusion by fitting the major axis velocity and velocity dispersion data of each galaxy (Figure 1) with a three component mass model, consisting of the sum of a central compact (spheroidal) stellar bulge, an exponential gaseous and stellar disk ($n_{Sersic}=1$), and a Navarro-Frenk-White (NFW) dark matter halo[22]. The output of the fitting is the dark matter to total mass fraction at $R_{1/2}$, $f_{DM}(R=R_{1/2})$ (Methods). We list the fitting results (and ±2 rms uncertainties) in Table 1 and summarize in Figure 3, which compares the high-redshift data to previous low-redshift results[7], and to the results of reference 6. With these basic assumptions we find that the ***dark matter fractions*** near the half-light radius for all our galaxies are ***modest to negligible***, even if the various parameter correlations and uncertainties are included (Figure 3 and Methods). We note that spatially variable $\sigma_0$ and deviations from planarity and exponential disk distributions undoubtedly make reality more complex than can be captured in these simple models.

All six disks are 'maximal'[7]. Their dark matter fractions are at the lower tail of local star-forming disks, and in the same region of $v_c$-$f_{DM}$ parameter space as local massive, passive galaxies[23] and some strongly bulged, early-type disks[7]. Dark matter fractions drop with increasing redshift from 0.8-2.3[6] (Table 1, Figure 3). The agreement of the dark matter fraction in the redshift=0.6-2.6 star-forming galaxies and local passive galaxies is interesting. Passive galaxies are likely the descendants of the massive 'main-sequence' star-forming population we are observing in our IFS-samples. Their star formation was likely quenched rapidly at redshift≤2 once they had grown to $M_{*Schechter}\sim 10^{10.6\ldots 10.9}\ M_{\odot}$, after which they transited to the passive galaxy sequence[24]. The low dark matter fractions in the high-redshift star-forming galaxies may thus be preserved in the 'archaeological' evidence of the local passive population.

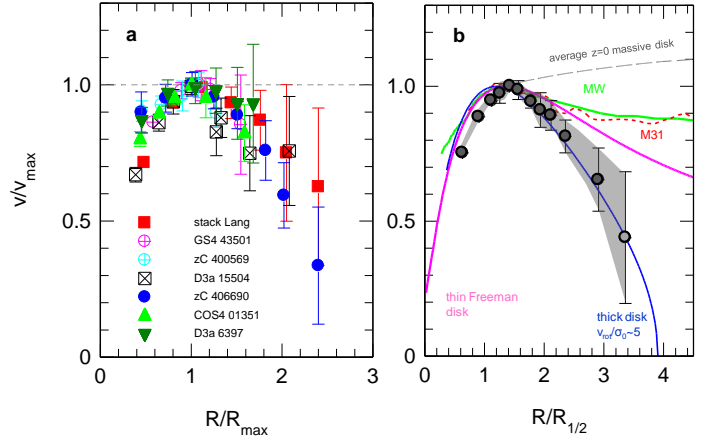

**Figure 2. Normalized rotation curves. (a):** The various symbols denote the folded and binned rotation curve data for the six galaxies in Figure 1, combined with the stacked rotation curve of 97 z=0.6-2.6 star-forming galaxies[18] (Methods). For all rotation curves we averaged data points located symmetrically on either side of the dynamical centres, and plot the rotation velocities and radii normalized to their maximum values. Error bars are ±1 rms. (**b**): The black data points denote the binned averages of the six individual galaxies, as well as the stack shown in (**a**), with 1 rms uncertainties of the error-weighted means shown as grey shading (with the outermost point in lighter shading to indicate that only two data points entered the average). In individual galaxies the observed value of the radius of maximum rotation amplitude, $R_{max}$, depends on the ratio of bulge to total baryonic mass of the galaxy, the size of the galaxy and the instrumental resolution, leading to varying amounts of beam smearing. We assume an average value of $R_{max}\sim 1.3$-$1.5\ R_{1/2}$. For comparison, the grey dashed line denotes a typical rotation curve slope of low-z massive, star-forming disk galaxies[19] (log($M_*/M_\odot$)~11), comparable to the six galaxies in our sample; the dotted red curve is the rotation curve of M31, the Andromeda galaxy[17], and the green curve that of the Milky Way[20]. The thick magenta curve is the rotation curve of an infinitely thin 'Freeman' exponential disk[21] with $n_{Sersic}=1$. The blue curve is a turbulent, thick exponential disk, including 'asymmetric drift' corrections for an assumed radially constant velocity dispersion of 50 km/s (and a ratio of rotation velocity to dispersion of 5)[17].

**Table 1. Physical Parameters of Observed Star-Forming Galaxies**

|  | COS4 01351 | D3a 6397 | GS4 43501 | zC 406690 | zC 400569 | D3a 15504 |
|---|---|---|---|---|---|---|
| redshift | 0.854 | 1.500 | 1.613 | 2.196 | 2.242 | 2.383 |
| kpc/arcsec | 7.68 | 8.46 | 8.47 | 8.26 | 8.23 | 8.14 |
| **Priors:** | | | | | | |
| $M_*$ ($10^{11}$ $M_\odot$) | 0.54±0.16 | 1.2±0.37 | 0.41±0.12 | 0.42±0.12 | 1.2±0.37 | 1.1±0.34 |
| $M_{baryon}$(gas+stars) ($10^{11}$ $M_\odot$) | 0.9±0.5 | 2.3±1.1 | 0.75±0.37 | 1.4±0.7 | 2.5±1.2 | 2.0±1.0 |
| H-band $R_{1/2}$ (kpc) | 8.6±1.3 | 5.9±0.8 | 4.9±0.7 | 5.5±1 | 4±2 | 6.3±1 |
| inclination (°) | 75±5 | 30±5 | 62±5 | 25±12 | 45±10 | 34±5 |
| dark matter concentration parameter c | 6.8 | 5 | 5 | 4 | 4 | 4 |
| **Fit parameters**: | | | | | | |
| $v_c(R_{1/2})$ (km/s)[a] | 276 | 310 | 257 | 301 | 364 | 299 |
| $R_{1/2}(n=1)$ (kpc) | 7.3 | 7.4 | 4.9 | 5.5 | 3.3 | 6 |
| $\sigma_0$ (km/s) | 39 | 73 | 39 | 74 | 34 | 76 |
| $M_{baryon}$(gas+stars, including bulge) ($10^{11}$ $M_\odot$) | 1.7 | 2.3 | 1.0 | 1.7 | 1.7 | 2.1 |
| $M_{bulge}/M_{baryon}$ | 0.2 | 0.35 | 0.4 | 0.6 | 0.37 | 0.15 |
| $f_{DM}(R_{1/2})=(v_{DM}/v_c)^2\|_{R=R1/2}$[b] | 0.21 (±0.1) | 0.17 (<0.38) | 0.19 (±0.09) | 0.0 (<0.08) | 0.0 (<0.07) | 0.12 (<0.26) |

[a] Total circular velocity at the half-light radius (rest-frame optical) $R_{1/2}$, including bulge, exponential disk (n=1) and dark matter, and corrected for asymmetric drift: $v_c(R)^2 = v_{rot}(R)^2 + 3.36\,\sigma_0^2 \times (R/R_{1/2})|_{n=1}$.
[b] Ratio of dark matter to total mass at the half-light radius of the optical light, $f_{DM}(R_{1/2})=(v_{DM}/v_c)^2_{R=R1/2}$, with numbers in the parentheses giving the ±2 rms ($\delta\chi^2=4$, ~95% probability) uncertainties, or upper limits. We use an NFW halo of concentration parameter c, and no adiabatic contraction.

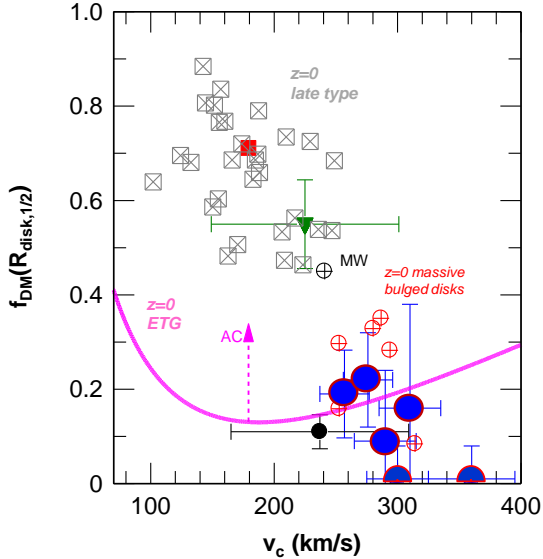

**Figure 3. Dark matter fractions.** Dark matter fractions from different methods are listed as a function of the circular velocity of the disk, at the half mass/light radius of the disk, for galaxies in the current Universe and ~10 Gyr ago. The large blue circles with red outlines indicate the dark matter fractions derived from the outer-disk rotation curves of the six high-z disks presented in this paper (Table 1), along with the ±2 rms uncertainties of the inferred dark matter fractions and circular velocities. The black circle and the green triangle denote the average dark matter fractions (and their ±1rms errors/dispersions in the two coordinates) obtained from the comparison of inner rotation curves and the sum of stellar and gas masses for 92 z=2-2.6 and 106 z=0.6-1.1 star-forming galaxies, respectively[6]. We compare these high-z results to z=0 estimates obtained from different independent techniques for late-type, star-forming disks (crossed grey squares, red filled square), for the Milky Way[20] (open crossed black circle,), massive bulged ($M_{*,bulge} \sim M_{*,disk}$), lensed disks[7] (crossed red circles), and for passive early-type galaxies[23] (thick magenta line). The upward magenta arrow marks the typical change if the z=0 data dark matter halos are maximally adiabatically contracted.

Why should high-redshift disks have been more baryon dominated than low-redshift disks? For one, they were gas rich and compact. Redshift~2.3 star-forming galaxies had ~25 times larger molecular gas-to-stellar mass ratios[25] ($M_{molgas}/M_* \sim (1+z)^{2.7}$), 2.4 times smaller sizes[26] ($R_{1/2} \sim (1+z)^{-0.75}$), and thus more than two orders of magnitude greater molecular gas surface densities than local galaxies. Large gas columns can easily dissipate angular momentum and drive gas inward. The strong redshift dependence of the gas fractions could explain the drop of $f_{DM}$ between redshift=2.3 and 1 (Figure 3). Massive galaxies at high redshift are thought to grow by rapid gas accretion, gas rich mergers and star formation triggered by this accretion[27-28]. In this dissipation dominated peak phase of galaxy growth the *centres* of dark matter halos can become baryon-dominated in 'compaction events' triggered by mergers, disk instabilities or colliding streams in the intergalactic web[29], or by gas 'pile-up' at early times when the gas accretion rates were larger than the star formation consumption rates. However, abundance matching results indicate that the *average* stellar-to-dark-matter ratio at the virial radius of the halo does not depend strongly on cosmic epoch and is well below the cosmic baryon fraction, $f_b \sim 0.17$, suggestive of very efficient removal of baryons due to galactic outflows[30].

Could the baryon dominance be caused by a lack of dark matter in the inner disk? These dark matter haloes could deviate from the standard NFW profile, with low concentration parameter (c<5), if they were still growing rapidly and not yet in equilibrium, or if they were perturbed by strong stellar and AGN feedback. We briefly discuss some of these alternatives, such as low concentration parameters, in Methods, where we also give a first quantitative comparison between observations and simulations.

Acknowledgments. We would like to thank our colleagues at ESO-Garching and ESO-Paranal, as well as in the 3D-HST and SINFONI/zC-SINF and KMOS/KMOS$^{3D}$ teams for their support and high quality work, which made these technically difficult observations possible. DW and MF acknowledge the support provided by DFG Projects WI 3871/1-1 and WI 3871/1-2. JC acknowledges the support of the Deutsche Zentrum für Luft- und Raumfahrt (DLR) via Project ID 50OR1513. TA and AS acknowledge support by the I-CORE Program of the PBC and Israel Science Foundation (Center No. 1829/12). We thank Simon Lilly and Avishai Dekel for valuable comments on the manuscript.

Authors Contributions:
Drafting Text, Figures, Methods Section: RG, NMFS, HÜ, TN, LT, OG, DL, AR, RS, AS,
Data Analysis and Modeling: RG, HÜ, PL, SW, RD
Data Acquisition and Reduction: RG, NMFS, HÜ, PL, LT, EW, SW, AB, SB, JC, MF, AG, JTM, KT
KMOS$^{3D}$ and SINS/zC-SINF IFS Survey Design and Management: NMFS, RB, EW, CMC, AR, RS, ST, DW
3D-HST Survey Analysis: NMFS, SW, GB, IM, EJN
Theoretical Interpretation: TN, TA, AB, SG, OG

The authors have no competing financial interests.

Correspondence and requests for materials should be addressed to R.Genzel (genzel@mpe.mpg.de) and/or Natascha Förster Schreiber (forster@mpe.mpg.de ).

# Methods

## Galaxy Samples

The galaxies discussed in this paper were taken from two near-IR integral field spectroscopic samples ('IFS-samples') of distant, massive ($\log(M_*/M_\odot) \gtrsim 9.6$) star-forming galaxies, targeting primarily rest-frame optical emission around the H$\alpha$ line,

- The SINS/zC-SINF survey of $z \sim 1.5 - 2.5$ star-forming galaxies[3, 31] observed with SINFONI on the ESO Very Large Telescope[32, 33] (VLT) in both seeing-limited mode ($2R_{1/2,beam}$ = FWHM ~ 0.45″-0.6″) and at higher resolution with adaptive optics (AO; $FWHM$ ~ 0.15-0.25″).

- The first 2.5-year sample from the ongoing five-year KMOS$^{3D}$ survey of $0.6 < z < 2.6$ galaxies[8], all observed in seeing-limited mode ($FWHM$ ~ 0.4″-0.7″) with the KMOS multiplexed IFS instrument on the VLT[34].

The IFS-sample galaxies are bench-marked to be representative in all important physical parameters of the high-z, 'field' star-forming population (stellar mass, star formation rate, size). We have not included galaxies in very dense ('cluster'-like) environments, and mergers or strongly interacting galaxies are relatively rare[8]. These galaxies are located on and around (±0.6 dex) of the 'star-formation main sequence' at all redshifts included. Below, we summarize the selection criteria used for drawing the individual objects and the stacking sample from the SINS/zC-SINF and KMOS$^{3D}$ surveys, and then discuss how their stellar, structural, and kinematic properties were derived. A comprehensive description of the surveys, target selections, and galaxy properties can be found in the references listed at the end of this Methods section.

### *Outer Rotation Curve Samples*

The six individual galaxies studied here were selected for very sensitive follow-up observations primarily based on their having (i) high quality and high signal-to-noise data in the initial SINFONI and KMOS observations, (ii) rotation-dominated disk kinematics from H$\alpha$, with a ratio of rotation velocity to velocity dispersion of $v_{rot}/\sigma_0 > 3$, (iii) no bright neighboring galaxy in line or continuum emission, and (iv) extended well-resolved star-forming disks. All six galaxies lie close to the main-sequence of star-forming galaxies at $\log(M_*/M_\odot) \geq 10.5$. By design, their half-light radii ($R_{1/2} \gtrsim 4$ kpc) place them in the upper half of the size distribution of massive star-forming galaxies, such that their disk emission probes further into the halo than for typical star-forming galaxies and their inferred baryonic angular momenta are comparable or larger than the population average[35] (~ 0.037). Smaller galaxies would therefore place less stringent constraints on the dark matter fraction in the outer disk, while they are likely to be even more baryon dominated[4]. While an optimum selection of rotation curve candidates would obviously favor high inclination galaxies, the six galaxies exhibit a wide range of inclination ($i \sim 25° - 75°$), where the low inclination systems are drawn from the adaptive optics SINS/zC-SINF sample (Förster Schreiber et al. in prep.), which is of high-quality in terms of angular resolution and depth but too small to allow an inclination cut.

For the stacking analysis[18], we selected star-forming galaxies based on their SINFONI or KMOS H$\alpha$ kinematics and their *Hubble Space Telescope* (*HST*) rest-optical morphologies as follows: (i) rotation-dominated disk kinematics from H$\alpha$ with $v_{rot}/\sigma_0 > 1$, and (ii) detection of a significant change of slope in the extracted velocity curve. The latter criterion is the most stringent one: while roughly 3/4 of the parent KMOS$^{3D}$ and SINS/zC-SINF samples are fairly unperturbed, rotation-dominated disks, only half of them show a flattening of the velocity curve at large radii. This large downsizing from the initial samples is driven by the limited field of view (FOV) of the IFS data not reaching the radius of velocity turnover for the more extended galaxies (~ 3″ × 3″ for the KMOS IFSs with which the largest parent sample was observed and for the AO-assisted SINFONI data with a 0.05″ pixel$^{-1}$ scale, and ~ 4″ × 4″ for the seeing-limited SINFONI data at 0.125″ pixel$^{-1}$ with the on-source dithering strategy used for most targets), or insufficient signal-to-noise at large radii to detect a change of slope. The resulting stacking sample consists of 103 near-main-sequence galaxies spanning $0.6 \leq z \leq 2.6$ and $9.3 \leq \log(M_*/M_\odot) \leq 11.5$. Their size and angular momentum distributions overlap well with those of the bulk of star-forming galaxies in similar redshift and mass ranges; there is only a mild bias towards larger $R_{1/2}$ for galaxies entering the stack especially for the $z > 1.3$ subset, caused primarily by the implicit signal-to-noise requirement at large radii imposed by the necessity of detecting a velocity curve flattening. We performed simulations to determine that the selection criteria and analysis methodology (see below) did not introduce a significant bias in the shape of the resulting stacked rotation curve. A more detailed discussion of the stacking methodology etc. is contained below

Extended Data Figure 1 compares the distributions in stellar mass, star formation rate, and size properties of the six individual galaxies, the stacking sample, and the parent SINS/zC-SINF and KMOS[3D] samples to those of the underlying $\log(M_*/M_\odot) \geq 9.0$ galaxy population drawn from the 3D-HST survey source catalog at $JH_{AB} <$ 26 mag (for which the mass completeness is below $\log(M_*/M_\odot) = 9.0$ out to z = 3)[13, 36] in the five CANDELS extragalactic fields[9, 10]. To account for the significant evolution in star formation rate and size at fixed mass, the Figure shows the offsets in star formation rate relative to the main sequence at the same z and $M_*$[37], and the offsets in effective radius at rest-frame 5000 Å relative to the mass-size relation for star-forming galaxies[26].

## Parent kinematic samples

The SINS/zC-SINF Hα sample was drawn from large optical spectroscopic surveys of high-redshift candidates photometrically pre-selected in multi-band imaging surveys based on various magnitude and/or color criteria at optical to mid-infrared wavelengths. As extensively described in two articles[3, 31], the SINS/zC-SINF Hα sample collectively provides a reasonable representation of massive star-forming galaxies at $z \sim 1.5 - 2.5$, with some bias towards bluer systems stemming from the primary criterion of having a secure optical spectroscopic redshift. In addition, the objects chosen for the near-infrared SINFONI observations were also required to have Hα falling away from bright OH sky lines and within high atmospheric transmission windows, and to have an expected total Hα flux of $\gtrsim 5 \times 10^{-17}$ erg s$^{-1}$ cm$^{-2}$ or, correspondingly, a star formation rate of $\gtrsim 10$ M$_\odot$ yr$^{-1}$ assuming a typical visual extinction of $A_V \approx 1$ mag. The latter flux criterion was applied last and did not significantly impact the final set of targets.

The KMOS[3D] sample is taken from the "3D-HST" source catalog[12, 13, 36] in fields with deep HST imaging from the CANDELS Treasury program[9, 10] and $R \sim 130$ $\lambda = 1.1 - 1.7$ μm slitless spectroscopy using the HST WFC3/G141 grism from the 3D-HST Treasury program[39]. Multi-band photometry and grism spectra are extracted for sources reaching $JH_{AB} \sim 26$ mag, redshifts are determined from the grism spectra (probing Balmer/4000 Å continuum break and/or emission lines) combined with the full optical-to-mid-infrared photometry and supplemented with spectroscopic redshifts available from the literature[13,36]. The KMOS[3D] targets are selected primarily to have a stellar mass $\gtrsim 10^{9.5}$ M$_\odot$ (derived from modeling the photometry; see below) and $K_{AB} = 23$ mag, and a sufficiently accurate redshift (dominated by grism and spectroscopic redshifts) ensuring avoidance of sky lines and poor telluric transmission for Hα. No explicit cut in star formation rate, color, or size is applied. Both the 3D-HST and KMOS[3D] near-infrared and mass selections result in a wide coverage of the full underlying galaxy population $0.7 < z < 2.7$, including the redder dust-obscured and/or more passive objects. In the first 2.5 years of our on-going KMOS[3D] survey, we emphasized high-mass targets (but over a very wide range of three orders of magnitude in specific star formation rate relative to the "main sequence" relationship). The resulting mass distribution is weighted towards high masses compared to a purely mass-selected sample; once accounting for this bias, the sample provides however a good representation of the underlying galaxy population in other properties[6, 8, 35].

## Global stellar and gas properties

The global stellar properties were derived following the procedures outlined by reference 40. In brief, stellar masses were obtained from fitting the observed broadband optical to near-/mid-IR (rest-UV to optical/near-IR) spectral energy distributions (SEDs) with population synthesis models[41], adopting a reddening law[42], an IMF[43], a solar metallicity, and a range of star formation histories (constant star formation rate, exponentially declining or increasing star formation rates with varying e-folding timescales). Over the mass and redshift ranges of the galaxies, gas-phase O/H abundances inferred from rest-optical nebular emission lines suggest metallicities of $\sim 1/4$ to $\sim 1\times$ solar[44-51]. Varying the assumed metallicity in this range would change the stellar masses in our modeling by < 0.1 dex[3, 52]. Given this small impact and the uncertainties in metallicity determinations for high-z star-forming galaxies[53], we chose to keep a fixed solar metallicity. We note that throughout the paper, we define stellar mass as the 'observed' mass ('live' stars plus remnants), after mass loss from stars.

The star formation rates were obtained from rest-frame ultraviolet + infrared luminosities through the Herschel-Spitzer-calibrated ladder of star formation rate indicators[40] or, if infrared luminosities are not available (from lack of observations or the source was undetected), from the broadband SED modeling described above.

Individual determinations of molecular gas masses (from CO line or submillimeter/far-infrared dust continuum emission) are scarce for our galaxy samples, and atomic hydrogen masses are not known for any of our high-z star-forming galaxies. We computed molecular gas masses from the general scaling relations between star formation rates, stellar masses, and molecular gas masses for main sequence galaxies (as a function of redshift)[25, 54]. We assumed, as argued previously[25], that at z ~ 1-3 the cold gas content of star-forming galaxies is dominated by the molecular component such that the atomic fraction can be neglected. As such the gas masses estimated from these scaling relations may be lower limits.

### *Structural properties*

The stellar structural parameters used as priors in our detailed kinematic modeling were derived by fitting two-dimensional Sérsic models to the high-resolution *H*-band images from *HST* observations available for all KMOS[3D] objects and most SINS/zC-SINF galaxies[6, 26, 55-60]. For the remainder of the SINS/zC-SINF galaxies (without *HST* imaging), structural parameters were estimated from the line integrated Hα distributions and/or from the continuum images synthesized from the IFS data, depending on the galaxy. To first order this approach is justified as high-redshift star-forming galaxies are gas-rich with large star formation rates and young stellar populations although there are some small systematic differences on average because of the presence of substantial stellar bulges in the more massive star-forming galaxies and of possible bright clumps and asymmetric distributions in tracers of on-going star formation[58, 61-65]. Among the individual galaxies and stacking sample of interest in the present study, only D3a 6397 lacks *HST* imaging.

For all but the most massive z ~ 1 – 3 star-forming galaxies, the stellar and Hα surface brightness distributions of main-sequence galaxies across the mass- and redshift range discussed in this paper are reasonably well fit by near-exponential (Sérsic index $n_S$~1-1.5) profiles[58, 62, 64-69] (also Förster Schreiber et al., in preparation). Above $\log(M_*/M_\odot)$ ~ 11, z ~ 1 – 3 star-forming galaxies feature a prominent stellar bulge component with characteristic $R_e$ ~ 1 kpc and median bulge-to-total ratio in *H*-band light reaching ~ 20% – 30% (and higher ~ 40% – 50% in terms of stellar mass ratio[58, 64, 65]). While central drops in Hα equivalent width, and even flux, of individual galaxies appear to be more frequent at the high-mass end, there is nonetheless evidence that on average the profiles are near-exponential[70] and that otherwise large central gas and dust concentrations may be present and obscure the optical light from star-forming regions in the inner parts of massive star-forming galaxies[63, 71-73]. These findings motivate our modeling assumption that the baryonic component is distributed in a compact spheroidal bulge at the center of an exponential disk.

The half-light (effective) radii $R_{1/2}$ estimates used for the galaxies refer to the major axis radii. To determine the inclination from the best-fit axial ratios, we account for the fact that for the mass range $\log(M_*/M_\odot)$>10 spanned by most galaxies of interest here, half or more of the high-redshift star-forming galaxies consist of symmetric, oblate thick disks[38, 74, 75], linked to their large intrinsic gas velocity dispersion[3, 8, 76-79] and $v_{rot}/\sigma_0 \lesssim 10$. We note that the *H* band probes a significant range of rest-frame wavelengths over the 0.7 < z < 2.7 of the individual galaxies and stacked sample considered in this work. A "k-correction"[26] should ideally be applied to the structural parameters for consistency (and similarly, a statistical correction between Hα and rest-optical sizes could be applied)[70]. However, the uncertainties adopted for the size and inclination priors are significant and larger than these corrections would be, and our modeling procedure accounts for them.

### *Kinematic classification and properties*

The kinematic information was derived from the SINFONI and KMOS data cubes following our well-established methods[3, 8, 80] (where all details of the data reduction and calibration procedures can also be found). In summary, for each galaxy we fitted Gaussian line profiles to each IFS spatial pixel in the final reduced data cube, in some cases after some prior smoothing to increase signal-to-noise ratios. The main kinematic parameters of interest are then derived from the resulting spatially-resolved maps of the velocity centroid and velocity dispersion: $v_{rot}$, $\sigma_0$, and $PA_{kin}$. The quantity $v_{rot}$ is the maximum rotational velocity corrected for beam smearing and inclination *i* ($v_{rot} = c_{psf,v} \times v_{obs}/\sin(i)$), where $v_{obs}$ is half of the difference between the maximum positive and negative velocities on both sides of the galaxy, and $c_{psf,v}$ is the beam smearing correction for velocity. The quantity $\sigma_0$ is the intrinsic velocity dispersion, corrected for beam smearing ($\sigma_0 = c_{psf,\sigma} \times \sigma_{obs}$), where $\sigma_{obs}$ is the measured line width in the outer parts of the galaxy *corrected for instrumental spectral resolution* (i.e., subtracting in quadrature $\sigma_{instr}$), and $c_{psf,\sigma}$ is the beam smearing correction for the velocity dispersion. $PA_{kin}$ is the position angle of the kinematic major axis passing through the extrema of the velocity field (the '*line of nodes*'). The beam smearing corrections were derived based on model disks for a range of masses, inclinations, ratios of galaxy to beam size, and radii of measurement appropriate for the SINS/zC-SINF and KMOS[3D] data sets; all details are given in Appendix A of reference 35).

With these kinematic maps and properties, a galaxy is classified as a "rotation dominated" disk if [8]

- the velocity map exhibits a continuous velocity gradient along a single axis; in larger systems with good signal to noise ratio this is synonymous with the detection of a 'spider' diagram in the two-dimensional, first moment velocity map[81];
- $v_{rot}/\sigma_0 > 1$
- the position of the steepest velocity gradient, as defined by the midpoint between the velocity extrema along the kinematic axis, is coincident within the uncertainties with the peak of the velocity dispersion map;
- the morphological and kinematic major axes are in agreement (≤ 30 degrees); and
- the kinematic center of the galaxy coincides with the maximum/centroid of the stellar distribution.

For the seeing limited KMOS$^{3D}$ survey, 83% of the resolved galaxies fulfill criteria 1 and 2 (92% at $z \sim 1$ and 74% at $z \sim 2$)[4]. This fraction slowly drops if the stricter criteria 3, 4 and 5 are added, and amounts to 70% if all 5 criteria are used. Similar results are obtained in the other recent surveys[62, 65, 78] (also Förster Schreiber et al. in preparation, for the SINS/zC-SINF sample).

The six individual galaxies discussed in this paper, and the stacking sample, satisfy all of the above disk criteria based on their SINFONI and KMOS data. With the higher signal-to-noise of the deeper follow-up observations of the individual galaxies, additional asymmetric features became apparent in the kinematic maps of three of them, associated with gas inflows towards the center (D3a 15504)[76], and gas outflows from the inner few kpc plausibly driven by an AGN and from bright off-center clumps driven by star formation (D3a 15504, zC 406690, zC 400569, D3a 6397)[82-84].

## Kinematic Analysis and Mass Modelling

In our kinematic analysis we proceed as follows below. For details we refer the reader to published papers[14, 20, 35], which provide more details on the mathematical and fitting methodology. These papers also refer to the publicly available data analysis tools in the ESO KMOS analysis pipeline as well as the LINEFIT tool[80] and a generalized fitting tool, MPEFIT, as well as the general cube analysis tool QFitsView (http://www.mpe.mpg.de/~ott/QFitsView/) to fit the kinematics by a combination of an exponential disk and dark-matter halos.

- We fit a Gaussian profile to each spaxel of the data cube, suitably smoothed to deliver sufficiently high SNR in the outer parts of the galaxy. We infer the systemic redshift (or velocity) of the galaxy by symmetrizing the red- and blue-shifted peak velocities. We determine the position angle of the kinematic axis $PA_{kin}$ by determining the line of nodes along the maximum velocity gradient, as well as the morphological major axis of the galaxy $PA_{morph}$ from the stellar surface density map or the H-band HST map, and take an average. These two angles typically agree to better than 5-7 degrees. Finally we determine the galaxy centre $x_0,y_0$ from an average of the zero-crossing of the line of nodes, the velocity dispersion peak and with the position of the central stellar bulge, which is prominent in all 6 of our galaxies. This allows the determination of $x_0,y_0$ to about 0.5-1 pixel, which is small compared to the size of the final extraction slitlet in the next step. We infer the disk's inclination from the minor to major axis ratio, $q=b/a$, of the stellar distribution or the HST H-band data, and $\cos^2(i) = (q^2 - q_0^2)/(1 - q_0^2)$, with $q_0 \sim 0.15 - 0.2$ appropriate for $z \sim 1 - 3$ (see references above). We have found in KMOS$^{3D}$ that this method works reasonably well for inclined disks[6] but naturally becomes very uncertain for the three face-on disks in our sample, D3a 15504, zC 406690 and D3a 6397. In these cases we included the inclination as a second order fit parameter to bring the disk mass into a good match with the prior $M_{baryon}=M_{bulge}+M_*(disk)+M_{gas}(disk)$.

- Next we go back to the original data cube and extract position-velocity and position-velocity dispersion cuts along the best fit major axis by fitting Gaussian profiles in software slits typically four to five pixels perpendicular to, and two to three pixels along the line of nodes. This corresponds to about 1.1 and 0.5 resolution elements, respectively, and thus is approximately Nyquist sampled along the line of nodes. To include systematic uncertainties, we multiply all fit errors by 1.5 and introduce lower ceiling uncertainties of ±5 and ±10 km/s for velocity and velocity dispersion measurements, respectively. These scalings are derived from bootstrapping, as well as error scaling of the final reduced chi-squared to about 1 for five of the six sources. We do not consider the additional information from deviations from Gaussian shape (the $h_3$ and $h_4$ components)[85], as this does not add any useful information on the outer disks, and only on the details of the heavily beam-smeared data in the cores of the galaxies, which is of secondary interest in this paper. We next subtract in squares the instrumental line width to infer intrinsic velocity dispersions in each slit.

- Now we begin the fitting process to a model that consists of an n=1 exponential disk (with effective radius $R_{1/2}$(disk)), a central bulge of $R_{bulge}$=1 kpc (with the B/T fit parameter giving the mass in the bulge relative to the total (disk+bulge) baryonic mass $M_{baryon}$(disk), the rest being in the n=1 disk), plus a dark matter halo. In those cases where we have a significant bulge, we assume that the Hα light distribution only traces the disk component, which is empirically justified[61, 63]. The assumption of exponential disk distributions (n=1) is supported quite well for the 3D HST star-forming population, which is characterized by n=0.5-2 for most but the more bulge dominated systems[6, 58]. We assume that the disk has a constant velocity dispersion, which we determine mainly in the outer parts of each galaxy where the beam-smeared rotation component is small. For the fitting we use a thin-exponential disk rotation curve[21] but correct for the effects of the significant scale height ($h_z \sim (\sigma_0/v_{rot}(R_{1/2})) \times R_{1/2}$) with models[86], which assume a constant scale height. Near $R_{1/2}$ the Freeman disk rotates 15% faster than the corresponding spherical model, and the Noordermeer correction is about 0.9-0.97 so that the final rotation velocity is only 4-11% faster than the spherical model. For the dark matter halo, we use an NFW[22] model. We do not consider adiabatic contraction of the halo, such that the only parameter of the halo is its mass $M_{virial}=M_{200(NFW)}$, where we fix the concentration parameter $c$ to a typical value at the respective redshift[87-89]. The angular momentum parameter of the halo is implicitly contained in $R_{1/2}$(disk) and the assumption $j_{disk}=j_{DM}$[35]. This yields 5 or 6 primary fit parameters, $R_{1/2}$(disk), $M_{baryon}$(disk), $M_{virial}$, B/T, $\sigma_0$ and in three cases, also the disk inclination. As for the data we determine velocity and velocity dispersion cuts for the model, convolve with the instrumental beam and then find the weighted best fit in the usual $\chi^2$ minimization (using the DYSMAL tool)[6, 80]. For four or five of these parameters we have priors from independent data: Inclination, B/T and $R_{1/2}$ from the HST J, H imagery, $M_{baryon}$(disk) from the sum of stellar disk mass in bulge and disk from the 3D-HST modelling and the (molecular) gas mass from the scaling relations between $M_*$, star formation rate and $z$ and $M_{molgas}$.[25, 54] We use these priors and their uncertainties to set upper and lower limits for the fitting range, as well as a constraint for inclination, as described above. The dark matter halo mass $M_{virial}$ is a free parameter in the fitting, with only a fixed lower limit of $M_{virial}=10^7$ M$_\odot$.

- We include in our fitting a correction of the disk rotation for pressure effects due to the significant turbulent motions (asymmetric drift)[6, 14, 35]. This correction lowers the rotation velocity in the outer disk significantly,

$$v_{rot}^2 = v_{circ}^2 + 2\sigma^2 \times \frac{d\ln\Sigma}{d\ln R}$$
$$= v_{circ}^2 - 2\sigma^2 \times \left(\frac{R}{R_d}\right) \quad (1).$$

Here $\Sigma$ is the surface density distribution of the disk, which we assume to be an exponential with scale length $R_d=R_{1/2}/1.68$. The assumption of radially constant velocity dispersion likely is a simplification. In that case the thickness of the disk increases exponentially with radius. A superposition of a thin and thick disk could lead to a radially increasing dispersion and a more steeply falling rotation curve. This would create room for a somewhat higher dark matter fraction at $R_{1/2}$ and beyond than in the case of constant dispersion. In absence of any clear evidence we prefer to stay with the simplest assumption of a constant velocity dispersion.

Extended Data Figure 2 shows the final $\chi_r^2$ distributions for the most important parameter for our study, the dark matter fraction at $\sim R_{1/2}$, $f_{DM}(R_{1/2})$. The number of independent data points in the six galaxies varies between 30 and 54, and with the error-scaling described above the best fit in five of the 6 galaxies has a minimum of $\chi_r^2$(min)=0.98-1.7. The one exception, D3a 15504 (black line in Extended Data Figure 2) has $\chi_r^2$(min)=2.25. This is caused mainly by the overshooting of the central velocity dispersion in two independent data sets that cannot be matched by the bulge plus disk data, as well as a second outlier point in velocity just north of the nucleus. We believe that the cause for these large near-nuclear deviations from the model is strong non-circular, or bar streaming in combination with outflows from the central AGN, which are well known to exist in this galaxy[76, 84]. We will come briefly back to this issue when we discuss the two-dimensional residual maps below.

### *Parameter correlations and Dark Matter fractions*

From Extended Data Figure 2 it becomes clear that the $\chi_r^2$ vs $f_{DM}(R_{1/2})$ space is relatively flat for some of our galaxies, and it is therefore important to systematically test the dependence of $f_{DM}(R_{1/2})$ on physical properties which constrain the rotation curve other than $M_{baryon}$ and the mass of the dark matter halo $M_{DM}$, namely $R_{1/2}$ and $B/T$. For galaxies with very low $f_{DM}(R_{1/2})$ (zC 400569 and zC 406690), varying $R_{1/2}$ within the uncertainties, or $B/T$ in steps of *0.05* or *0.1*, does not significantly alter $f_{DM}(R_{1/2})$. For the other galaxies, changing $B/T$ has a larger effect than changing $R_{1/2}$. We find most extreme changes for galaxy D3a 6397, with $(\partial f_{DM})/(\partial B/T)=2.4$ when going from the best-fit *B/T=0.35* to *B/T=0.3* (i.e. towards lower $f_{DM}(R_{1/2})$), and with $(\partial f_{DM})/(\partial R_{1/2})=0.6$ when going from the best-fit *$R_{1/2}$=0.87* to *$R_{1/2}$=0.97* (i.e. towards higher $f_{DM}(R_{1/2})$). We show schematically in Extended Data Figure 3 how $f_{DM}$ and $\chi^2$ change for changes in these various parameters.

Generally, we find that increasing $B/T$ leads to increased $f_{DM}(R_{1/2})$, which can be understood in the sense that a higher fraction of the baryonic mass in the central bulge decreases the relative contribution of the baryons to $v_{circ}$ at $R_{1/2}$. For changing $R_{1/2}$, the effects are less definite, but for the majority of cases we find that increasing $R_{1/2}$ again leads to increased $f_{DM}(R_{1/2})$. For these cases, this can be understood in the sense that a larger $R_{1/2}$ distributes the baryonic mass onto a larger disk (i.e. less compact), leading to less relative contribution of the baryons to $v_{circ}$ at $R_{1/2}$. Mean changes in $f_{DM}(R_{1/2})$ when increasing/decreasing the best-fit $R_{1/2}$ by the uncertainties given in Table 1, or increasing/decreasing the best-fit $B/T$ by 0.1, are all below $\Delta f_{DM}$ =0.13 (see also Extended Data Figure 3).

The dark matter fraction depends also on the halo mass distribution. Our best-fit models do not include the possible adiabatic contraction of the dark matter halo as a response to the formation of the central galaxy. Simulations show that at high redshift (*z~2*) adiabatic contraction can have an effect on the central dark matter density distribution[90, 91]. Its net effect, however, is not well constrained and depends largely on the feedback implementation[92], where strong supernova or AGN feedback can even create central dark matter cores[93, 94]. Recent work[95] constructs a toy model of repeated inflow and outflow cycles for dwarf to Milky Way type haloes to capture the various effects of baryons on the central dark matter distribution. They find that high gas fractions and low star formation efficiencies favour halo expansion, as well as extended baryon distributions. Another possibility is 'baryonic gas-pile-up' at early times[103]. Since the star formation accretion rate scales as $\sim(1+z)^{2...3}$ (103) while the star formation efficiency scales as $(1+z)^{0.6}$ (25), star formation may not be efficient enough at z>2-3 to consume the incoming accreted baryonic gas, and gas might pile-up in the inner disk[103].

We tested the effect of adiabatic contraction on an NFW halo[96]. Generally, for the constraints on the dark matter halo as set by our data, we find that the effects of adiabatic contraction (or of modest variations of the concentration parameter) on $f_{DM}$ are lower than the effects of changing $R_{1/2}$ or $B/T$ within the uncertainties (Extended Data Figure 4).

## Two-dimensional Analysis

Our analysis so far has used major axis cuts of velocity and velocity dispersion to characterize the mass distribution. We used the availability of the unique two dimensionality of integral field data to constrain $x_0,y_0$ and $PA_{kin}$. The question is whether two dimensional fitting of the gas kinematics might provide additional constraints. It turns out that for the relatively low resolution data on compact high-z disks (with $R_{1/2}(disk)/R_{beam}$~2.7-4.5) indeed most of the rotation curve information is encapsulated on/near the major axis. In deep AO data on moderate inclination, large disks, a kinematic estimate of inclination may be obtained from the off-axis data, in addition to centre and node direction. The 22h SINFONI AO data for D3a 15504 (Förster Schreiber et al. in preparation) can be used in this way, and Extended Data Figure 5 shows the result. The analysis of the minor axis cut data in this case support the evidence from the HST stellar distribution and the baryonic to dynamical mass constraint that the inclination of this galaxy is low, $34°\pm5°$ (Table 1).

### *Kinematic residual maps*

Another way of utilizing the full two-dimensional information is to construct two-dimension velocity and velocity dispersion residual maps (data minus model), constructed from the major-axis cut method, and then check whether systematic residuals appear[81]. Extended Data Figure 6 shows the corresponding residual maps for all 6 galaxies.

The residual maps broadly show that the simple model of a compact bulge, plus n=1 thick disk, plus NFW halo model does a fairly good job in accounting for the data. With two or three exceptions discussed below, most of the residual maps do not show large scale features or strong deviations that are comparable to the amplitudes in the original maps. Average values in the twelve residual maps range between -10 and +10 km/s, a few percent of the maximum data range. The median dispersions of the residuals range from 8-18 km/s, comparable to the measurement errors in most individual pixels, with two exceptions.

### Kinematic anomalies: D3a 15504

One outlier is the galaxy D3a 15504 (top left in Extended Data Figure 6). As already discussed in the first paper discussing this galaxy[76], the strong (±65 km/s), velocity gradient near the nucleus but along the minor axis of the galaxy had been apparent, which the deeper data confirm. Related to this minor axis streaming, plus probably also influenced by a strong, broad nuclear outflow component[84], is the high value of the nuclear velocity dispersion (~175 km/s), which cannot be accounted for by the best-fitting disk model and thus shows up as a large outlier in the velocity dispersion residual map. D3a 15504 has a small neighbour ~1.5" NW of the nucleus (PA= -45$^0$, visible in the stellar density contour map in Figure 1), of mass 2-3x10$^9$ M$_\odot$, ~3% of the stellar mass of the main galaxy. Hα emission from the satellite is detected in our deep integral-field data at a projected velocity of ~+10 km/s (relative to the systemic velocity of the main galaxy), ~140 km/s redshifted relative to the projected rotation velocity of the main galaxy at this radius. The neighbour thus is a satellite. The position-velocity diagram of Hα emission shows that the satellite is connected back to the main galaxy, clearly indicating that the two are interacting. We have removed the well separated Hα emission of the satellite before fitting the rotation curve in Figure 1.

### Kinematic anomalies: zC 406690

zC 406690, the second galaxy from the top in Extended Data Figure 6, exhibits a significant anomaly in both velocity and velocity dispersion residual map as well, in the outer south-western part of the rotating ring structure, on and near 'clump B' [82, 83]. This anomaly is caused by localized blue-shifted, very broad (up to -1000 km/s) Hα emission near that clump, and can probably be explained by star formation driven outflows. zC 406690 also has a companion located about 1.6" W of the main ring galaxy, with 6x10$^9$ M$_\odot$ in stellar mass (14% of the main system). There is a marginal detection of Hα from that companion at ~+100 km/s relative to the systemic velocity of the main galaxy, in which case it is an interacting satellite.

### Kinematic anomalies: zC 400569

zC 400569, the third galaxy from the top in Extended Data Figure 6, has two neighbours. Both are fairly prominent in Hα, but are not detected in [NII], plausibly because of their low mass and metallicity. The larger one, with stellar mass ~7x10$^9$ M$_\odot$ (5% of the mass of the main galaxy, 1.3x10$^{11}$M$_\odot$) is 1" to the south-east of the nucleus of zC 400569, with a projected velocity in Hα of -330 km/s, appears to be edge on but does not exhibit much of a velocity gradient along its major axis. The second neighbour, 1.5" to the south-east, has a stellar mass of 3x10$^9$ M$_\odot$ (2.3% of the main galaxy) and shows an east-west velocity gradient of ±20 km/s around the centre, which is blue-shifted by -410 km/s relative to the main galaxy. It thus appears that zC 400569 is situated in a group of gas rich, satellites galaxies of low mass and metallicity. The effect of the first galaxy on the line profiles is visible in the lower left of the velocity and velocity dispersion residuals in Extended Data Figure 6, but does not very much affects the dynamical analysis in Figure 1.

### Neighbours and Warps

Our analysis of the HST and Hα data of the six galaxies shows that five sample galaxies have neighbours projected within 1" and 3.3" (8-25 kpc) of the main galaxy, and between 1% (GS4 43501, 4x10$^8$ M$_\odot$) and 20% (COS4 01351, 1.2x10$^{10}$ M$_\odot$) of its mass. Of those five the detection of Hα emission from the companion in three (in D3a 15504, zC400690, zC400569) shows that these companions are indeed interacting satellites but probably not in the other two galaxies (COS4 01351, GS4 43501).

The Jacobi (or Roche, or Hills) radius in a double galaxy system with mass $M_1$ and $M_2$, separated by $R_{12}$, defines the distance from the lower mass system $M_2$, within which tidal forces by the smaller system strongly perturb particles in the bigger system. This radius is given by (**97,** Chapter 8)

$$r_J = \left(\frac{M_2}{3 \times M_1}\right)^{1/3} \times R_{12}$$
$$= 3 \text{ kpc} \left(\frac{M_2/M_1}{0.05}\right)^{1/3} \times \left(\frac{R_{12}}{12 \text{ kpc}}\right) \quad (2)$$

Here we have already inserted the typical mass ratios and separations for the satellite-main galaxy systems in our sample. This simple analysis shows that tidal perturbations or stripping by the satellite can be somewhat important in the outer parts of the main galaxies, if the satellite is within ~$R_{1/2}$ of the main galaxy, as seems to be the case, for instance, for the D3a 15504 north-western satellite.

In addition to interactions, warping can be important in the outer disks and is frequently observed in the outer HI layers of z~0 galaxies (including the Milky Way)[98]. Theoretically this type of buckling or firehose instability (with a predominant m=2 mode) can occur in galaxy disks with surface density Σ, with radial wavelengths λ less than

$$\lambda \leq \lambda_J = \frac{\sigma_x^2}{G\Sigma} \qquad (3),$$

where $\sigma_x$ is the in-plane velocity dispersion, (**97**, Chapter 6.6.1; **99**) if the system is sufficiently cold in the vertical direction for the instability to grow, which requires[99, 100]

$$h_z < \frac{\sigma_z^2}{G\Sigma}.$$

$$\text{such that } \frac{\sigma_z}{\sigma_x} < a_{crit} \sim 0.3....0.6 \qquad (4).$$

The current data for high-z galaxies suggest that the velocity dispersion ellipsoid is isotropic[4, 82], $\sigma_x = \sigma_z$, such that warping should be suppressed.

If the warp has a sufficiently high amplitude, it could indeed introduce a radial dependence of the peak rotation velocity along the major axis. If the dominant mode is uneven m=1 (or m=3, as in the Milky Way)[101], warps would also introduce the same sign of the change in the absolute value of the peak rotation curve on the blue- and the redshifted side of the galaxy, which could mimic a radial decrease (or increase) in the rotation curve, with equal probability. However, no significant *increase* is seen in any of the galaxies of the SINS/zC-SINF or KMOS[3D] samples. If the mode is even (m=2), one should observe rotation curves that increase on one side, and decrease on the other. We do not observe this feature in any of the six galaxies presented here. Finally, the phase of the warp does not have to be aligned with the major axis and might change with radius. Such precessions could be observed in the residual maps. The data do not show any evidence for such an effect.

## Comments on Overall Strategy

This paper puts forward observational evidence for very high baryon fractions (and correspondingly low dark matter fractions) in several high mass disk galaxies at z~1-2. First hints for this result came from Hα kinematics studies initially in the SINS sample[3], and more robustly in the KMOS[3D] sample[6, 8] and, independently by others in compact high-z star-forming galaxies[4], as well as in the MOSDEF survey[5]. In these cases one infers the dynamical mass of the inner star-forming disks of z~1-2.5 star-forming galaxies from the peak Hα rotation velocity (or velocity dispersion), and then compares it with the baryonic mass, that is, the sum of the stellar mass and the cold gas mass. The stellar mass is estimated from population synthesis fitting of the UV/optical SEDs, and the gas mass from CO or dust tracing the molecular hydrogen content[3-6, 25]. Unfortunately this '*inner disk dynamics*' technique, even in the best cases of spatially well resolved kinematics, requires strong assumptions on star formation histories, initial stellar mass functions and CO/dust emission to $H_2$-mass calibrations that are inherently not known better than ±0.2 to ±0.25 dex. Close to 50 years of experience of local Universe studies have taught that robust statements on dark matter content cannot be done solely, and certainly not robustly, from the 'central disk dynamics' technique[102].

More robust statements on dark matter fractions are expected to come from '*rotation curves in the outer disk and inner halo*'. We detected a few cases of potentially falling Hα rotation curves in the best SINS/zC-SINF data a few years ago but it took several years to collect enough integration time to make a solid case for the six galaxies reported in this paper. However, these cases are obviously biased from the outset as we invested additional integration time only in galaxies for which we already had prior evidence of falling rotation curves. In order to check that these galaxies are not 'outliers' in a population of star-forming galaxies with primarily flat-rotation curves, we needed a statistical statement on the occurrence of falling rotation curves in the high-z, massive star-forming population, even if the individual rotation curves in these other galaxies were individually not good enough for study. As the KMOS[3D] sample grew in size, we thus developed as a third element of our strategy a '*rotation curve stacking*' technique on the overall sample to test for the hypothesis that the individual cases were (or were not) outliers. The results of this project[18] do confirm that falling rotation curves appear to be common at z~1-2.5, and are referred to in this paper (and briefly summarized in the next section) but an exhaustive technical discussion of the methodology, required to convince readers of the robustness of the result, did not fit into the current Nature Letter and had to be placed in an independent paper[18].

A fourth approach in this overall strategy is the measurement of the baryonic and stellar mass, Tully-Fisher relation, zero point offsets as a function of redshift. This '*TF-zero point evolution*' approach is somewhat related to the 'central disk dynamics' method above, but does not rely on the accurate determination of baryon fractions in individual galaxies but the redshift dependence of a population averaged property (the TF zero-point). The results of the TF-technique for the KMOS[3D] sample will be reported in a forthcoming paper (H.Übler et al., in preparation) and are in agreement with the three other tests.

## The Stacking Analysis of Lang et al. [18]

In the following we summarize the stacking analysis in the paper by (**18**) and refer the reader to that paper for a more detailed description.

Having established the properties of a few high quality outer disk rotation curves in star-forming galaxies with long integrations, the next step is to characterize the *average* rotation curve of a *representative sample* of z~0.6-2.6 massive star-forming disks, as drawn from the seeing-limited KMOS$^{3D}$ and AO-assisted SINS/zC-SINF datasets. For this purpose (**18**) employed a stacking approach to systematically determine the shape of outer rotation curves.

This stacking method is designed to leverage the faint outer ionized gas emission combining the signal of >100 massive star-forming galaxies at 0.6 < z < 2.6. The methodology of this stacking technique first includes the normalization of each individual rotation curve by its observed maximum velocity $V_{max}$ and the corresponding turnover-radius $R_{max}$. Both $V_{max}$ and $R_{max}$ are determined by fitting the rotation curve with an exponential disk model convolved with the appropriate instrumental resolution of the data set. Values for $R_{max}$ are also independently derived for each galaxy by converting intrinsic half-light radii measured on rest-frame optical HST images into observed turnover-radii, also taking into account the effect of beam smearing as well as the shape of the mass distribution (as parametrized by the Sersic index). These independently derived $R_{max}$ are in good agreement with the $R_{max}$ values measured on the actual rotation curves and thus substantiate the validity of our $R_{max}$ measurements using pure exponential disk models. Based on mock galaxy simulations, the paper[18] demonstrates that the above technique of normalizing and stacking rotation curves is able to reproduce both outer falling and rising rotation curves, with the latter being expected in case massive star-forming galaxies at high redshift are genuinely more dark-matter dominated.

With $R_{max}$ and $V_{max}$ derived for each galaxy, normalized position-velocity diagrams are generated which are then averaged into a stack from which a final combined rotation is curve is constructed. Due to field-of-view limitations for both KMOS and SINFONI AO observations, the number of galaxies available at a given galacto-centric radius drops with distance to the centre, such that the combined stacked rotation curve can be reliably determined out to ~2.4 $R_{max}$, corresponding to several effective radii. Within this radius, the shape of the resulting stack outlines a fall-off in rotation velocity beyond $R_{max}$, symmetric on either side from the center, reaching down to ~ 0.65 $V_{max}$. We show in the left panel of Figure 2 a slightly altered version of the original stack[18], where we remove D3a 15504, zC 400569 and GS4 43501, *so that the remaining stack and the individual rotation curves are completely independent of each other.* The study[18] utilizes template rotation curves of local spiral galaxies[19] and show that the outer fall-off in the stacked rotation curve deviates significantly from the (mildly rising) average rotation curves of local analogues of similar mass at the same galactocentric radii.

In addition, the study[18] evaluates the outer drop in their stack by a comparison with models including baryons arranged in exponential disk configuration with added dark-matter NFW halos, taking into account pressure gradients in the outer disk resulting from a significant level of velocity dispersion. This comparison demonstrates that the stacked rotation curves can be explained by high baryonic disk mass fractions ($m_d \geq$ 0.05), in combination with a significant level of pressure support in the outer disk. The latter is accounted for by considering a value 4.8 < $v_{rot}/\sigma_0$ < 6.3 depending on radius, as found to be the average for the sample of stacked galaxies. Considering galaxies with strong pressure support (represented by a low $V_{rot}/\sigma_0$) for stacking, the resulting averaged outer rotation curve steepens compared to a stack made with only high $V_{rot}/\sigma_0$ galaxies, which supports the conclusion that a significant part of the outer fall-off in the stacked rotation curve is driven by the presence of pressure effects in the outer disk.

The stacking study[18] furthermore demonstrates that the above results are largely independent of underlying model assumptions such as the presence or absence of a central stellar bulge, the halo concentration parameter *c*, and the possible adiabatic contraction of the host halo, since those do not appear to alter the shape of the expected rotation curve significantly.

The above results are in good agreement with the conclusion drawn from our six individual rotation curves. Most importantly, the stacking work[18] confirms that outer falling rotation curves are a common feature among a larger representative sample of massive star-forming galaxies at high redshift.

## Comparison to Simulations

In the following, we discuss how our results compared to state-of-the-art cosmological simulations of galaxy formation, and we briefly comment on the 'thick disk'-phenomenon in local spiral galaxies.

Any quantitative comparison between observations and simulations is challenged by the fact that fundamental properties of high-z galaxies are still not matched by the large-volume simulations that are needed in order to produce the observed diverse galaxy populations, as well as disk galaxies that are massive already at z~2. Also the effective co-moving spatial resolutions in current simulations, such as Illustris and Eagle, are 1-3 kpc, which are not or only barely sufficient to resolve bulges, giant star-forming clumps and other intra-galactic structures, and the observed high star formation rates (SFR) and gas fractions are under-predicted in these simulations[104, 105]. Having said that, peaked rotation curves are produced in current simulations mostly as a result of weak (or no) feedback[106, 107]. However, it is consensus now that stronger feedback descriptions (momentum feedback from supernovae, or from active galactic nuclei) are needed in order to match many other observed galaxy properties, like outflows, disk-like morphology, or angular momentum[108-117].

Given the limitations outlined above, a one-to-one comparison of our results to simulations is not feasible. To nevertheless perform a qualitative comparison to our z~2 galaxies, we utilize results from the Illustris cosmological hydro-simulation as follows: we select galaxies at z=2 with stellar masses of $10.9<\log(M_*/M_\odot)<11.3$, and SFR>55 $M_\odot$/yr, a total of 80 galaxies. Their mean properties in terms of stellar size, stellar mass, and SFR are $<R_{1/2,*}>$ = 5.6 kpc, $<SFR>$=120 $M_\odot$/yr, $<M_*>$ = $1.2 \times 10^{11}$ $M_\odot$. We then inspect their two-dimensional distributions of velocity and velocity dispersion, where a cut of star formation rate surface density, $\Sigma_{SFR}>0.01$ $M_\odot$/yr/kpc$^2$, has been applied. Galaxies are classified as rotationally supported if they exhibit a continuous velocity gradient along a single axis, and if the velocity dispersion map displays a central peak which coincides with the kinematic center of the galaxy. This if fulfilled for ~50% of the galaxies. Judging from the velocity and velocity dispersion maps, ~25% of the rotationally supported Illustris galaxies exhibit falling rotation curves, i.e. 10-15% of the parent sample of 80 z=2 SFGs.

In other simulations falling rotation curves are seen typically for very compact systems, which do not represent the main population of simulated (or observed) SFGs. Most simulations find larger $v_{rot}/\sigma_0$ than seen in the observed system, perhaps due to their still modest spatial resolutions not capturing all the necessary intra-galactic physics mentioned above. We note that typical intrinsic velocity dispersions of the Illustris sample discussed above are 30-40 km/s. A recent paper connects peaked rotation curves in zoom simulations at z=3 to higher bulge-to-total fractions[118]. A significant fraction of our high-z SFGs have indeed massive bulges. However, (18) find that the effect of a bulge on the shape of the outer rotation curve is negligible for our otherwise fairly extended SFGs. Although our observed galaxies are all star-forming and extended, these theoretical results give some support to our interpretation of the evolution of our galaxies with cosmic time: the massive, high-z SFGs are likely soon to be quenched, and will afterwards evolve on the passive sequence into local, massive ETGs[24, 119].

There is an intriguing similarity between the turbulent high-z disks and the 'thick disk'-phenomenon in local spiral galaxies[20]. The stellar population of the thick disk indicates a formation time of z>1-2 while thin disk formation started at z~1 and lasts until today. Thus, it is conceivable that our high redshift sample of disk galaxies shows the transition period from thick disk to thin disk formation.

# Extended Data

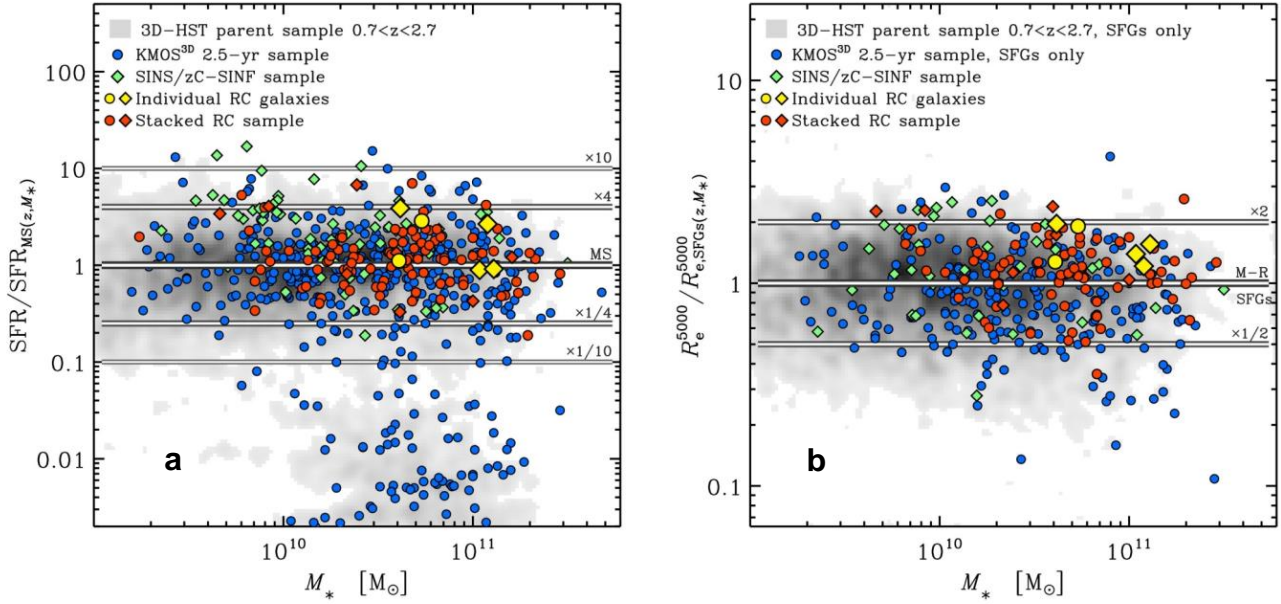

**Extended Data Fig 1. Location of the galaxies included in our analysis in stellar mass vs. star formation rate (a) and vs. size (b)**. In panel (**a**) the star formation rate is normalized to that of the "main sequence" [37] at the redshift and stellar mass of each galaxy. In panel (**b**) the size is the half-light radius measured in the observed *H*-band corrected to the rest-frame 5000 Å and normalized to that of the mass-size relation for star-forming galaxies[26] at the redshift and stellar mass of each source. In both panels, the greyscale image shows the distribution of the underlying galaxy population at 0.7<*z*<2.7 taken from the 3D-HST source catalog at $\log(M_*/M_\odot) > 9.0$ and $K_{AB} < 23$ mag (the magnitude cut applied when selecting KMOS$^{3D}$ targets and corresponding roughly to the completeness limits of the parent samples for SINS/zC-SINF targets). The current 2.5-year KMOS$^{3D}$ sample is shown with blue circles, and the SINS/zC-SINF sample, with green diamonds. The two KMOS$^{3D}$ and four SINS/zC-SINF galaxies with individual outer rotation curves are plotted as yellow circles and diamonds, respectively. Similarly, the KMOS$^{3D}$ and SINS/zC-SINF galaxies included in the stacked rotation curve are plotted as red circles and diamonds. All 3D-HST and KMOS$^{3D}$ galaxies are included in panel (**a**), while only star-forming galaxies are shown in panel (**b**), defined as having a specific star formation rate higher than the inverse of the Hubble time at their redshift. The individual outer rotation curve galaxies lie on and up to a factor of 4 above the main-sequence in star formation rate (with mean and median $\log(SFR/SFR_{MS}) = 0.24$ ), and have sizes $1.2 - 2$ times above the $M_* - R_e^{5000}$ relation (mean and median offset in $\log(R_e^{5000}) \approx 0.2$ dex). In star formation rate and $R_e^{5000}$, the distribution of the stacked rotation curve sample is essentially the same as the reference 3D-HST population in mean/median offsets ($\approx 0.06$ dex above the MS and $\approx 0.07$ dex above the mass-size relation) as well as in their scatter about the relationships ($\approx 0.3$ dex in $\log(SFR)$ and $\approx 0.17$ dex in $\log(R_e^{5000})$; see (**26, 37**).

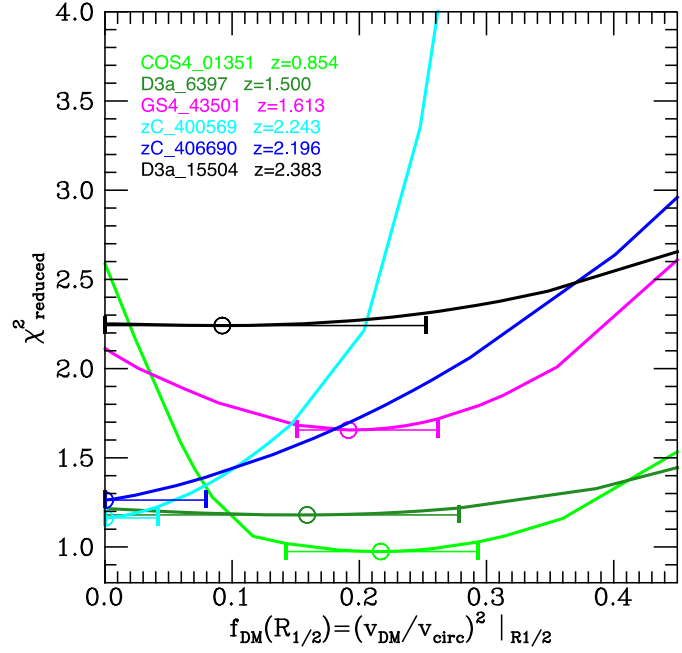

**Extended Data Figure 2. Quality of fit and error of parameter determinations.** The reduced chi-squared $\chi_r^2$ as a function of the dark matter fraction at $R_{1/2}$ for the six galaxies in our sample, once the other parameters ($x_0$, $y_0$, $PA_{maj}$, $i$, $\sigma_0$, $R_{1/2}$, B/T) are already fixed at their best fit values. Global minima are marked by a circle; error bars give $\Delta\chi^2=\pm 4$ ranges, corresponding to confidence levels of *95% (2 rms)* under the assumption of single-parameter Gaussian distributions. This is the most important parameter dependence for our data set.

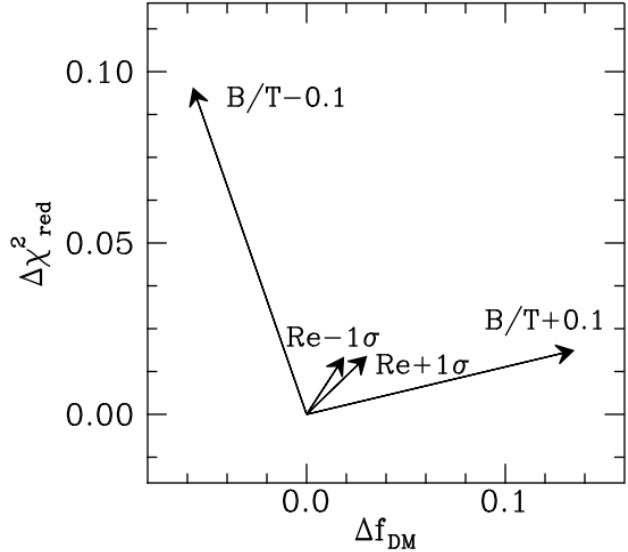

**Extended Data Figure 3.** Mean changes for COS 01351, D3a 6397, GS4 43501, D3a 15504 in $f_{DM}$ and $\chi_r^2$ for changes in the secondary parameters B/T and $R_e = R_{1/2}$.

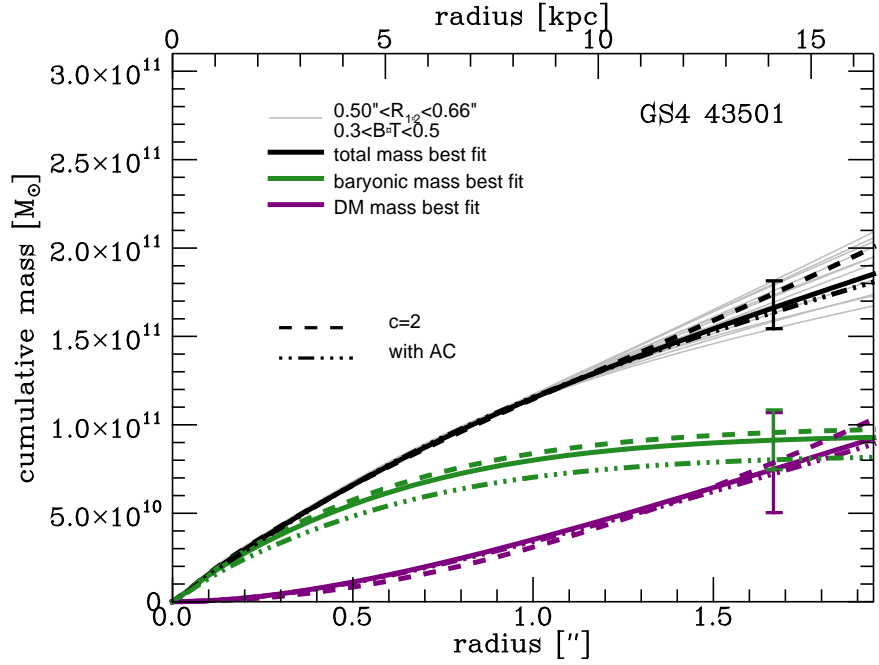

**Extended Data Figure 4. Cumulative mass as a function of radius for one of our model galaxies.** Solid lines show the best fit; error bars show the variations in total (black, grey), baryonic (green) and dark matter (purple) mass at the outermost projected radius constrained by our data, if deviations from *B/T* and $R_{1/2}$ within the uncertainties are considered (only cases with $\chi_r^2 < 1.75$ are considered). Dashed lines show the best fit for a model with lower concentration parameter (*c=2* instead of *c=5*), dashed-dotted lines show the best fit for a model with adiabatic contraction (**96**). Both modifications of the dark matter profile lead to changes of the cumulative mass which are smaller than those obtained through varying *B/T* and $R_{1/2}$ within the above uncertainties. The grey lines encompass variations of the dark matter fraction of $f_{DM}(R_{1/2})=[0.14;0.27]$ (best fit $f_{DM}(R_{1/2})=0.19$).

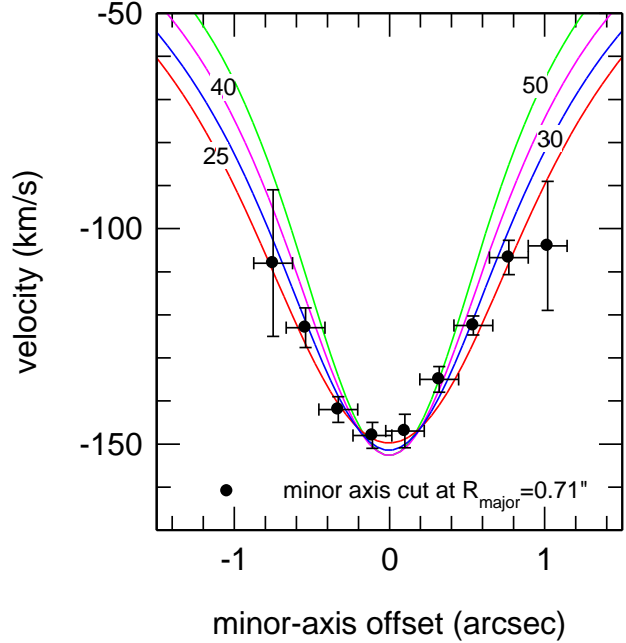

**Extended Data Figure 5. Minor axis cut at $R_{major}=0.71$" of D3a 15504.** Shown are the velocities along with disk models for different inclinations, 25° (red), 30° (blue), 40° (magenta) and 50° (green). The minor axis cut favours a low inclination. In combination with the morphology of the stellar surface density distribution (Figure 1) and the constraint on baryonic mass of the disk, this yields an overall inclination of 34°±5° (Table 1).

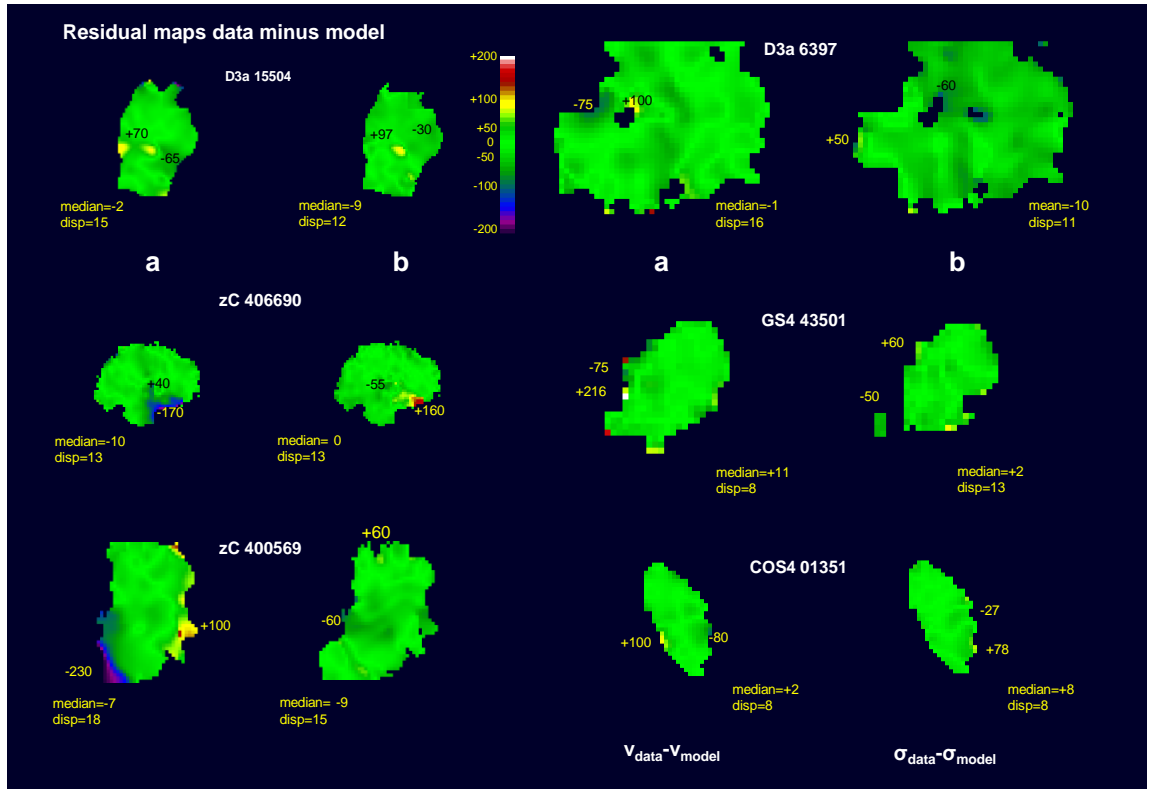

**Extended Data Figure 6. Velocity (a) and velocity dispersion residual maps (b, data minus model) for all 6 galaxies in this paper**. The color bar is the same in all maps (from -200 km/s (purple) to +200 km/s (white)). Minimum and maximum values are noted in each map, as are the median and median dispersion values.